\documentclass{mn2e}

\usepackage[dvips]{graphicx}

\begin{document}

\title[Star formation in Galaxy Pairs]
{Galaxy Pairs in the 2dF Survey I. Effects of Interactions in
 the Field}

\author[Lambas et al.]{Diego G. Lambas,$^{1,2,6}$,
  Patricia B. Tissera, $^{1,3}$,
M. Sol Alonso, $^{1,4}$ and
Georgina Coldwell, $^{1,2}$\\
$^1$ Consejo Nacional de Investigaciones Cient\'{\i}ficas
y T\'ecnicas.\\
$^2$ Observatorio Astron\'omico
de la Universidad Nacional de C\'ordoba,  Argentina.\\
$^{3}$ Instituto de Astronom\'{\i}a
y F\'{\i}sica del Espacio, Argentina.\\
$^{4}$ Complejo Astron\'omico El Leoncito, Argentina \\
$^6$ On a fellowship from Agencia C\'ordoba Ciencia \\
$^7$ John Simon Guggenheim Fellow.
}

\maketitle

\begin{abstract}
We study galaxy pairs in the field selected from the 100 K public release of
the 2dF galaxy redshift survey. Our analysis provides a  well
defined sample of 1258 galaxy pairs, a large database suitable for statistical
studies of galaxy interactions  in the local universe, $z \le 0.1$.
Galaxy pairs where selected by radial velocity ($\Delta V$) and projected
separation ($r_{\rm p}$)
criteria determined by analyzing the star formation activity
within  neighbours.
We have excluded pairs in high density regions
by removing galaxies in groups and clusters.
We analyze the star formation activity in the pairs as a function of both  relative
projected distance and relative radial velocity.
We found power-law relations for the mean star formation birth parameter and
equivalent widths of the
galaxies in pairs as a function of $r_p$ ad $\Delta V$.
We find
that star formation in galaxy pairs is significantly
enhanced over that of
isolated galaxies with similar redshifts in the field
for  $r_p < 25$ kpc and $\Delta V < 100$ km/s.
We detected that when compared to isolated galaxies of similar
luminosity and redshift distribution, the effects of having
a companion are more significant on the star formation
activity of bright galaxies in pairs, unless the pairs
are formed by similar luminosity galaxies. In this case,
the star formation is enhanced in both components.
The ratio between the fractions of star forming galaxies in pairs 
and in  isolation is a
useful tools to unveil the effects of having a close companion.
We found that about fifty percent of galaxy pairs do not show signs of
important star formation activity (independently of their
luminosities) supporting the hypothesis that the internal properties
of the galaxies play a crucial role in the triggering of star formation
by interactions.

\end{abstract}



\section{INTRODUCTION}

The  actual cosmology paradigm for galaxy formation
assumes that structure forms by hierarchical aggregation.
 In such scenarios
galaxy interactions are  frequent and have a crucial role in determining
galaxy properties.
The formation of galaxies in these schemes can be successfully analyzed
by using numerical simulations which show that, as the systems
are assembled, mergers and interactions can trigger star formation
with efficiencies that seem to depend mainly on the internal
structure of the systems.
By using pre-prepared mergers, Barnes \& Hernquist (1996) showed
 how the gas component experiences
torques originated in the companion, increasing its gas density
and triggering a starburst during the orbital decay phase of
the satellite. These starbursts are fed by
gas inflows  tidally induced if the axisymmetrical character
of the potential well is lost during the interaction.
The stability of the systems can be assured by a dominating
central mass concentration.  A second starburst could be generated
at the actual fusion of the baryonic cores.
 Cosmological hydrodynamical simulations showed that  these
processes take place in the formation of galactic systems in
a hierarchical aggregation scenario  in a similar way to that
shown by pre-prepared mergers.
Tissera (2000) found a correlation between mergers
and starbursts.
Recently, Tissera et al. (2002, Tis02) showed that
the effects of interactions are different at different stages of
evolution of the systems being  more efficient at higher $z$ when
the systems are in early stages of evolution, and consequently,
their potential well could be shallower.
If the structure in the Universe formed in consistency with
a hierarchical scenario, then, the proximity of  a companion could
affect the mass distributions and  trigger gas inflows
 producing an enhancement of the star formation (SF) activity.

In the local Universe, observations show that mergers
and interactions of galaxies
affect star formation in galaxies (e.g., Larson \& Tinsley 1978;
Donzelli \& Pastoriza 1997; Barton, Geller \& Kenyon 1998).
It has also been shown that the merger rates (e.g., Woods, Fahlman, Richer 1995; Le Fevre et al. 2001; Patton et al. 2002)
and the star formation activity of galaxies increase with
redshift, suggesting a change in the impact of
interactions on the SF process as galaxies evolve. Although the relevance of these violent events on the formation
of the structure and their SF history is now widely accepted, many questions
remain to be answered.
For example, as it has been
reported by other authors (e.g., Petrosian et al. 2002),
many interacting systems show weak star formation activity
suggesting that the particular internal conditions within these
systems  may be needed to trigger star formation.

An insight on the nature of interactions can be obtained from studies
of close pairs of galaxies in projection. Physical pairs must have
similar redshifts, where relative velocities affect distance
interpretation and therefore the
true relative separations.
Yee \& Ellingson (1995) and Patton et al. (1996) adopt a minimum projected
separation of $20 h^{-1} kpc$ to define close pairs and
they find no significant differences
between mean properties of paired and isolated galaxies although those
which appear to be undergoing interactions or mergers have
strong emission line and blue rest-frame colours.
A similar result by Zeff \& Koo (1989)
indicate that in some systems colours correspond to recently
enhanced star formation
with an overall  distribution consistent with  field galaxies.
Kennicutt et al. (1987) examined the $H_\alpha$ equivalent width, UBV colours and
far-infrared flux of a complete sample of local pairs of galaxies and found that
these close pairs exhibit a general trend of enhanced star formation and nuclear activity
but with a wide dispersion about the mean behavior.
Barton et al. (1998) analyzed a sample of approximately 250 pairs
of  galaxies
showing a correlation between their radial and velocity separations
and the $H_{\alpha}$ equivalent width.  Although these authors found
a clear correlation,
their pair sample is still small to carry out a
thoughtful statistical study of the effects of interactions and
their cosmological evolution.

The 100K public data release (Colles et al. 2001) of approximately 100000 galaxies
of the 2dF Galaxy Redshift Survey (2dFRS)
opens the possibility to study galaxy interactions by selecting
the largest galaxy pair catalog hitherto constructed
and analyzing their properties.
The 2dF public data provide the redshift, magnitude and
spectral types ($\eta$) of galaxies with $z \le 0.15$ (Madgwick et al. 2002a,
hereafter M02). Madgwick et al. (2002b) found a tight
correlation between $\eta$ and the birthrate parameter $b$.

In this work, we present a catalog of galaxy pairs in the field constructed
from the  2dF public release data and
 analyze their properties in comparison to isolated galaxies in the field.
For this purpose we will exclude those galaxies that belong
to groups as defined by Merch\'an \& Zandivarez (2001, hereafter MZ01).
 Pairs in high density environments will be analyzed in a forthcoming paper.
Here, we aim at answering questions such as which type of galaxies are
preferentially found in pairs,
how the star formation varies among them according to their luminosities, etc.
We also intend to statistically
determine a critical  relative radial velocity  and spatial separation
for the classification of galaxy pairs.
As it has been  pointed out in previous works, galaxy pairs
can be classified as interacting or close pairs. Interacting
galaxy pairs are those that show explicit signs of interactions such
as tidal tails. Close pairs are defined according to velocity difference and
spatial separation criteria (e.g., Barton et al.2002; Patton et al. 2002).
In this work, galaxy pairs will be statistically selected by
applying both a velocity and separation criteria which will
be determined according to the star formation  activity of neighbouring
galaxies.

Because the public release 2dF catalog
has different selection effect problems,
control samples sharing the same selection effects
will be constructed,  focusing the analysis
 on the statistical differences between
galaxies in pairs and isolated ones in the 2dFRS as a useful procedure
to unveil  the effects of galaxy interactions on the star formation
process.

In Section 2 we will defined the pair selection criteria ,
field galaxy pair catalog and the comparison samples.
Section 3 is focused on the analyses of the star formation properties
of galaxies in pairs and  Section 4 summarizes the main conclusions.

\section{GALAY PAIR CATALOG}

\begin{figure}
\includegraphics[width=84mm]{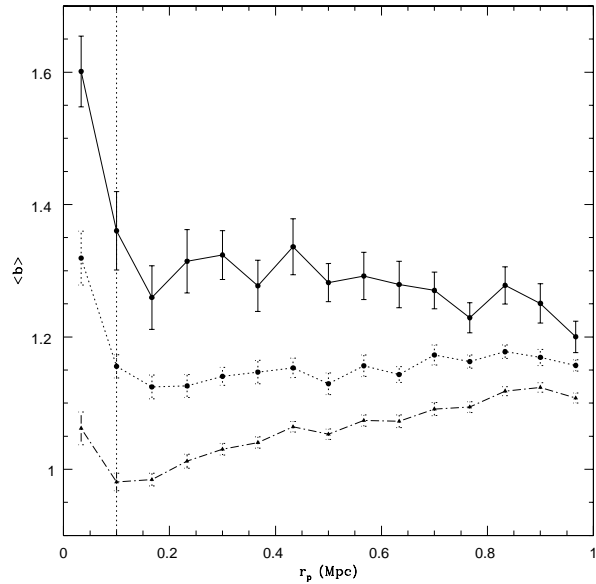}
\caption{Mean birthrate parameter $b$ as a function of relative
projected separation $r_p$
of galaxies with $\eta >3.5$ (solid line),
$\eta >-1.4$ (dotted line) and  no $\eta$ restriction (dotted-dashed line).
The dotted vertical line depict the spatial separation threshold
identification.
}
\label{brgeo}
\end{figure}

The 2dF Galaxy Redshift Survey public data consist of
102326  galaxies observed in the northern and southern main strips of
the final planned catalog which will have approximately 250000 galaxies.
Despite of the fact that the 2dF public catalog is not complete,
we argue that galaxy  pair searching is not
severely  affected by completeness effects.
In effect, in spite of the fact that the minimum fiber separation for
2dF spectroscopy is approximately 25 arcsec, the survey strategy was
to repeat the measurements in each field with new fiber positions
in order to achieve the highest completeness. Thus, from this point of view
there is no bias against small angular separations which would bias the
results specially at higher redshifts. Therefore, the inclusion of a pair
in our catalog depends mostly on the inclusion of each member in the survey,
which were randomly selected within the targets of each field. Therefore, we
conclude that there are not significant selection effects on the pair sample which
could bias our results on star formation activity.

The 2dF data comprise information on redshift ($z$), angular
separation, spectral type ($\eta$) and blue magnitude ($m_{b}$).

The spectral type  parameter $\eta$
are defined by a Principal Component Analysis, PCA, carried out
 by M02
where it was found to be  related to the morphological  type and the strength
of the absorption-emission features.
Madgwick et al. (2002a) suggest four different spectra groups:\\

\noindent  Type 1: $\eta< -1.4$\\
Type 2: $-1.4 \le \eta<1.1$ \\
Type 3: $1.1 \le \eta < 3.5$ \\
Type 4: $\eta \ge 3.5$.\\

 The  good correlation found between $\eta$
and EW($H_{\alpha}$) by M02 supports the interpretation of this parameter
as a good  indicator of star formation activity.

We  have estimated galaxy luminosities
 by applying the K-corrections derived by M02.
The public 2dF data have galaxies with redshift in the range up to  $z \approx 0.3 $,
 however,
for the purpose of defining galaxy pairs we restrict our analysis
to
 $z\leq 0.1$ in order to prevent the results for strong biases in
galaxy luminosities and unreliable  spectral type estimates
for distant galaxies.

Contamination by AGN activity could contribute to the emission spectral
features and therefore affect our interpretation of star formation
derived from the $\eta$ parameter. However, on the global sense
our conclusions should not be greatly affected
given the expected good correlation between
the star birth rate parameter $b$ and the $\eta$ parameter
in the models analysed by Madgwich et al. (2002b).

As we discussed in the introduction, Barton et al. (1998) presented
a  sample of approximately 500 galaxies in  pairs  which
were selected to have projected separations  $r_{\rm p} < 50 {\rm h^{-1}}
$ kpc and velocity separations $\Delta V \leq 1000 \ {\rm km/s}$.
With  this sample they studied the possibility that
 tidal interactions induced star formation.
However, it is not clear from their work, if there are  critical
spatial and   velocity separations which could  establish  limits for
tidal interactions to be  effective star formation triggering mechanisms.

We have estimated the stellar birthrate parameter,  $b=SFR/<SFR>$
for each galaxy  which provide a useful measure of  the
present level of star formation activity of a galaxy related to its
mean past SF history. This parameter has been found to correlate
with the morphological type (Kennicutt 1992) in the sense that
late-type spirals and starbursts have larger $b$ values.
According to the calibration shown by Kennicutt (1998), systems
undergoing strong star formation activity have $b >1$.

In order to compute the $b$ parameter we
assume a linear correlation between
$\eta$  with EW($H_{\alpha}$)  as reported
by M02: $EW(H_{\alpha})= 5.64\eta + 10.9$.
The final  relation between $b$ and $\eta$ was obtained by
fitting a linear regression  of the form
 $b = 0.045EW(H_{\alpha}) + 0.61$ to the
high star forming galaxies of  Carter et al. (2001).
The resulting  equation $b = 0.25  \eta + 1.06$  relates linearly
$b$ with $\eta$.
Our relation  reproduce very well the calibration
shown by Kennicutt (1998).
We notice that the dependence
of $b$ on $ \eta$ estimated by Madgwick et al. (2002b) is stronger than the
linear one adopted in this paper. It should be taken into account that
the birthrate parameter $b$ is linked to models for the star formation
history and so the results are reliable on a global or statistical
sense. Therefore, the conclusions obtained in this work are not expected
to depend crucially on the particular dependence of $b$ on $\eta$ as far as
$\eta$ provides a useful measure of star formation in galaxies.

For the purpose of analysing how $b$ depends on $r_{\rm p}$ and $\Delta V$
(defined as $\Delta V=c*(z_1-z_2)$,where $z_1$ and $z_2$ are the redshift
of the galaxies in the pair),
we  estimated the number of neighbours  within concentric spheres centred
at a given galaxy. Then, we calculated  the
 mean star formation rate parameter $<b>$
of these neighbours.
Three sub-samples were constructed according to the $\eta$ spectral
type of the galaxy centre.
 Sub-samples I,
II and  III
take as a centre a galaxy with $\eta >3.5$,
 $\eta >-1.4$ and  a any $\eta$ value, respectively.
We have adopted $\eta =-1.4$ to roughly segregate galaxies
between two main types: early and late ones, while
$\eta >3.5$ corresponds to systems undergoing strong star formation activity.
Sub-sample III has been constructed to  compare the two previous ones
 with the general 2dF survey.

All galaxies with $r_{\rm p} \leq 1$ Mpc and $\Delta V \leq 1000 \ {\rm km/s}$
were taken into account.
In Fig. \ref{brgeo}, for each sub-sample, we show the mean
birthrate parameter
in $r_{\rm p}$ bins.
Otherwise stated error bars in the figure have been
computed trough the bootstrap technique.
>From this figure we see that, when the central galaxy has
$\eta >-1.4$  or $\eta >3.5$,
  the $<b>$ parameter increases,
as the further away neighbours are  cast out.

 This result makes evident that the closer neighbours experience  the
stronger star formation activity.
This effect is more significant when the central galaxy is also experiencing
strong star formation activity ($\eta >3.5$).
We can also observed  that neighbours
of Type 4 galaxies within 1 Mpc have  $<b>$ larger than
the averaged one of 2dF galaxies at any $r_{\rm p}$.

\begin{figure}
\includegraphics[width=84mm]{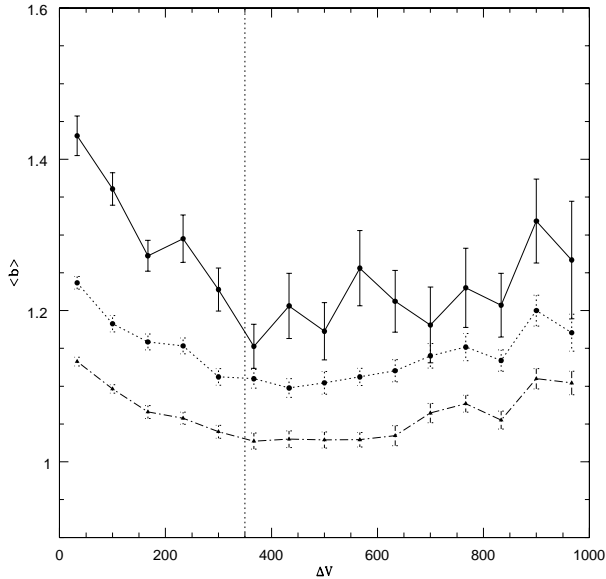}
\caption{Mean bithrate star formation parameter $b$ of galaxies
as a function of
relative velocity to the target:
$\eta >3.5$ (solid line),
$\eta >-1.4$ (dotted line) and no $\eta$ restriction (dotted-dashed line)
targets are shown.
The dotted vertical line depicts the relative radial velocity
threshold for galaxy pair identification.
}
\label{bvgeo}
\end{figure}

Similar calculations where done for  velocity bins within
$\Delta V < 1000 \ {\rm km/s}$. In Fig. \ref{bvgeo} we show
$<b>$ as a function of $\Delta V$ for  the  sub-samples constructed by
 the same criteria mentioned above.
We found a velocity cut-off of $350 \ {\rm km/s}$ (vertical dotted line), so
that galaxies with $\Delta V < 350 \ {\rm km/s} $  have $<b>$ significantly enhanced
with respect to the those   of the total 2dF survey  at a $3\sigma$ level.

Therefore,  according to this analysis,  $r_{\rm p}\leq 100$ kpc and
$\Delta V \leq 350 \ {\rm km/s}$ can be defined as reliable upper limits
for the   relative radial velocity  and projected distance criteria to
select galaxy pairs with  stronger specific star formation
than the averaged  galaxies in the 2dF. The signals are more
significant when the centre galaxies  belong to the spectral Type 4,
but it is statistically equivalent to that measured
 for  central  galaxies with $\eta >-1.4$.
Hence these values of relative projected separation and radial velocity
may be considered as suitable thresholds for the triggering of star
formation  induced by interactions.

\subsection{Field Galaxy Pair Catalog}

>From the previous analysis we
found that a projected distance of 100 kpc and
a  relative radial velocity  of 350 ${\rm km/s}$ are reliable limits
to select pairs with high probability of being an interacting
system  with enhanced
 specific star formation
activity. Galaxies situated at larger distances and with greater
velocity differences do not show statistically significant
signs of enhanced SF with respect to the background.
With these criteria, we constructed a  galaxy  pair (GP) catalog from the 2dF
public release data. And since,
  we are interested in studying those pairs
 that belong to the field (pairs in groups will be
discussed in a future paper), we cross-correlated
 the total galaxy pair
catalog with the 2dF group catalog developed by MZ02.
The resulting field galaxy pair  catalog (FGPC) comprises 1258 pairs with
$z \le 0.1$.

In order to properly assess the significance of the results obtained from
 the FGP catalog  we have
defined a sample of control galaxies from the 2dF catalog
by a Monte Carlo algorithm that selects for each galaxy pair in the FGP
catalog, two  galaxies in the
 field (i.e. not included in the FGP catalog and not members of the
MZ group catalog) with
the same redshift range of the galaxy pair with no restriction in their
relative distance.
The procedure followed to construct this control catalog
assures that
it will have the same selection  effects than the FGP catalog, and
consequently, it can be used to  estimate the actual difference
between galaxies in pairs and isolated galaxies, in the field, unveiling the
effect of the interactions.

For analyzing the spectral type composition of the
FGPs in comparison with the that of general galaxies in the field
we defined five combined categories ($C_i$) according
to the combination of spectral type of the galaxies in the pairs.
The first four correspond to GPs that have equal spectral
type:
$C1=(1,1)$, $C2=(2,2)$, $C3=(3,3)$ and $C4=(4,4)$.
The last spectral category corresponds to pairs composed of galaxies with
different spectral types (Table 1).

\begin{table}
\center
\caption{Spectral  composition of galaxy pairs:
Percentages of spectral type categories.}
\begin{tabular}{|c c c c c c }
Catalog & $C1$ & $C2$   & $C3$   & $C4$ & $C5$\\
FGP & 22 & 7 & 6 & 5& 59\\
CS & 7 & 10 & 6 & 3& 74\\
${\rm FGP/CS}$& 3.14 & 0.7 & 1 & 1.7 & 0.80
\end{tabular}

{\small Note: FGP: field galaxy pairs; CS: control
sample. }
\end{table}

\begin{figure}
\includegraphics[width=84mm]{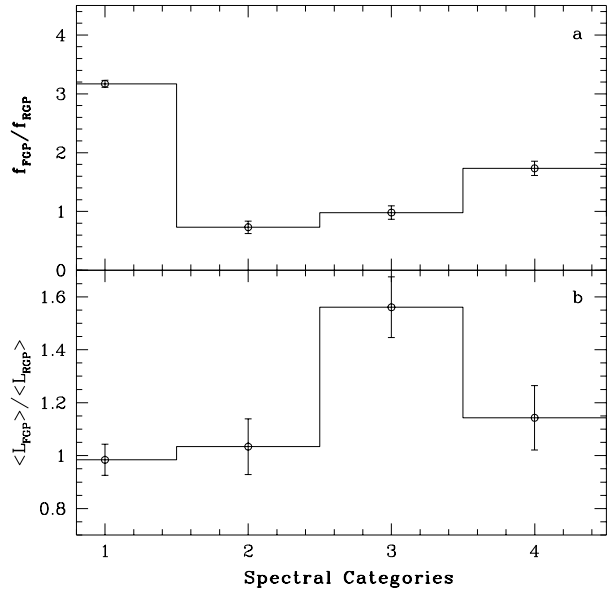}
\caption{a) Frequency of spectral type combinations of
field galaxy pairs (FGPs) and  b) Ratio between the mean luminosities of
galaxies in pairs and the corresponding to the control sample
as a function of  their spectral categories. In both cases,  we
have divided these relations by  the corresponding of  the control catalog.  
Error bars correspond to standard Poisson
deviations.
}
\label{esta-general}
\end{figure}

In Fig. ~\ref{esta-general}a we have plotted the ratio between the 
 fraction of FGPs
in the first four spectral categories   and of  those obtained from the
control sample ($F^{\star}=f^{\star}_{\rm FGP}/f^{\star}_{CS}$).
We see that there is excess of
C1 and C4 spectral category pairs with
respect to the control sample, suggesting a trend
for FGPs to be composed of
 two galaxies with similar spectral
characteristics, both non star forming or both
with significant star formation
(see Table 1).
We have estimated the mean luminosities of
$<(L_1 + L_2)/2>$,  of  each category pair and the corresponding to the control sample.
In Fig. ~\ref{esta-general}b
we show the corresponding ratios
of these two samples.
We can see that on average, galaxy pairs in C3 and C4 categories are
significantly more
luminous than their isolated counterparts.

\section{Star Formation in Galaxy Pairs}

The final  FGP catalog allows a  detailed study of
the possible effects of the interactions.
For these galaxies, we have estimated the  $<b>$ as a function of $r_{\rm p}$
and $\Delta {\rm V}$
where $r_{\rm p}$ and  $\Delta {\rm V}$ are the projected distance and
the  relative radial velocity , respectively, between members of a pair.

\begin{figure}
\includegraphics[width=84mm]{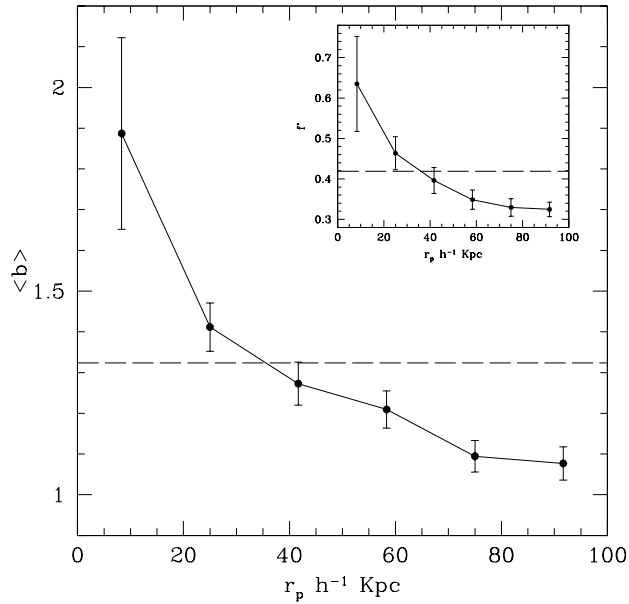}
\caption{Mean $b$ parameters estimated in projected distance
bins for galaxies in interacting pairs.
The dashed horizontal lines
represent the mean b parameter for the control sample.
The small box correspond to the fraction $f^\star$ of galaxies
with $b>\bar{b}$
}
\label{fig4A}
\end{figure}

\begin{figure}
\includegraphics[width=84mm]{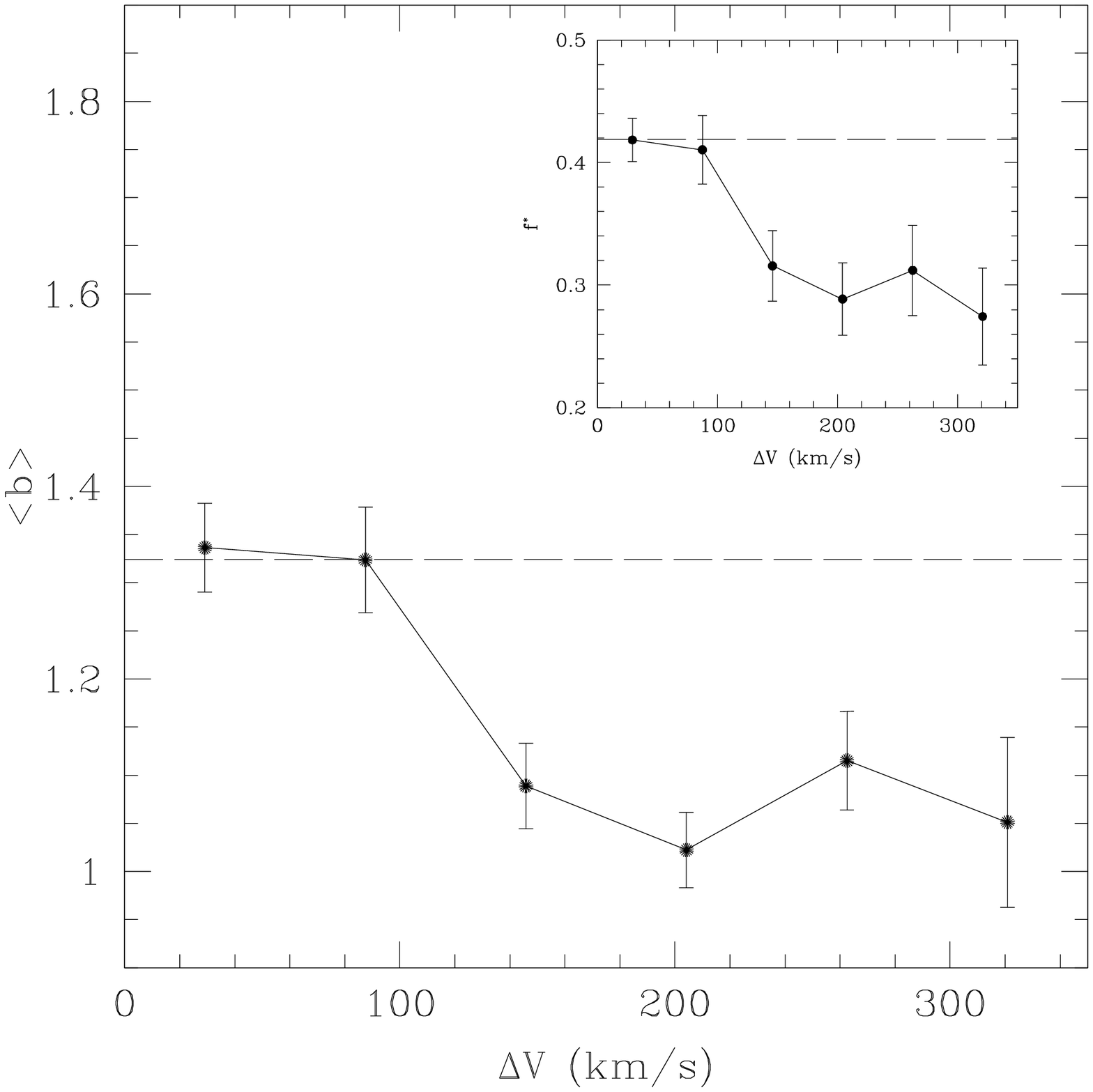}
\caption{Mean $b$ parameters estimated in relative radial velocity
bins for galaxies in interacting pairs.
The dashed horizontal lines
represent the mean b parameter for the control sample.
The small box correspond to the fraction $f^\star$ of galaxies
with $b>\bar{b}$
}
\label{fig4B}
\end{figure}

Fig. \ref{fig4A} shows $<b>$ in bins of  $r_{\rm p}$. It can be seen that
the star formation efficiency  increases
for  closer galaxy pairs. A similar behaviour is found
for the  relative radial velocity  (Fig. \ref{fig4B}).
Power laws  provide  good fits to  the data. The following relations
for the mean $b$ parameter and the equivalent width have been obtained
from these fittings:

\begin{equation}
<EW> = (4.10\pm + 0.50) r_{\rm p}^{-0.40 \pm 0.02}
\end{equation}

\begin{equation}
<EW>= (1.30 \pm 0.50) \Delta V^{-0.30 \pm 0.16}
\end{equation}

\begin{equation}
<b> = (0.63 \pm 0.01) r_{\rm p}^{-0.23 \pm 0.02}
\end{equation}

\begin{equation}
<b> = (0.40 \pm 0.31) \Delta V^{-0.14 \pm 0.09}
\end{equation}

where $r_{\rm p}$ and $\Delta V$ are given in Mpc and ${\rm km/s}$
and $EW$ in Amstrong.
Errors have been estimated by applying the bootstrap technique
(100 random samples).
Note that the dependence on the projected distance is
more significant than on the velocity separation.
These relations have been confronted with Barton et al. (1998) data
finding a good agreement with their observations for the
dependence on projected distance. 

In order to assess  the importance of interactions on  SF activity  compared
to isolated galaxies,
we have computed the mean
$b$ parameter for the galaxy pair control sample,  $\bar{b} = 1.32 $.
This value has been depicted in Fig. \ref{fig4A} and in Fig. \ref{fig4B}
(dashed lines)
where  it can be appreciated
that only galaxies
in very close pairs show significantly higher
mean star formation activity
than that of  isolated  galaxies in the
field.

Complications related to  the physical interpretation
of redshift space identified pairs where $r_{\rm p}$ and $\Delta V$
provide only lower limits for the true galaxy separations should
be always considered. Also, transient
properties, interlopers and interactions which have not undergone
a close approximation difficult a straightforward
interpretations of the mean values of the starbirth rate parameter.
Taking into account these caveats,
 we have  calculated the fraction $f^{\star}$ of star forming
galaxies with $b > \bar{b}$, and show them in
the small windows of Fig. \ref{fig4A} and in Fig. \ref{fig4B}.
The dashed lines represent the corresponding fractions for
the control sample.  As it can be appreciated
 the proximity in
 projected distance correlates with an increase
in the fractions of galaxies undergoing strong star formation
activity until they exceed the mean fraction of the control sample.
Note that although we find no excess of $b$ in FGP  with
respect to those of isolated ones as a function of the
relative radial velocity, there is a 
tendency in the fractions of pairs with $b > \bar{b}$ and
small $\Delta V$.
These results point out
the existence of a large number of pairs that, although
satisfying  the general constrains of velocity and spatial separations
do not exhibit enhanced SF with
respect to isolated galaxies.

The weaker dependence of the mean birth rate parameter $<b>$
on the relative velocity may have several sources.
Firstly,  an overall
rms uncertainty of $\simeq 85$ km/s is derived  by repeated observations in
the 2dF survey and by comparison with other redshift catalogs.
 This large scatter
implies a large observational error in the relative velocity of a pair of galaxies
in the catalog $\simeq 120$ km/s. Therefore,
this observational uncertainty should be considered as the main cause
of the weaker dependence on relative velocity. There are,
however, other issues to be considered, since
there can be several
pairs with large spatial separations but with
small radial velocity differences.
Hence, for a given relative velocity bin,  contributions of
pairs with different orbital position are canceled out.
This is reflected in the lack of large $b$ values for any 
relative velocity bin (notice however, that the global mean $b$ values
for all $r_p$ in Fig. 4, and all $\Delta V$ in Fig. 5  are equal).
In the case of projected separation bins, there is also a
contamination by small $r_{\rm p}$ and large $\Delta{V}$
although the enhancement of star formation with
spatial separation is strong enough to override
this contamination.

\begin{figure}
\includegraphics[width=84mm]{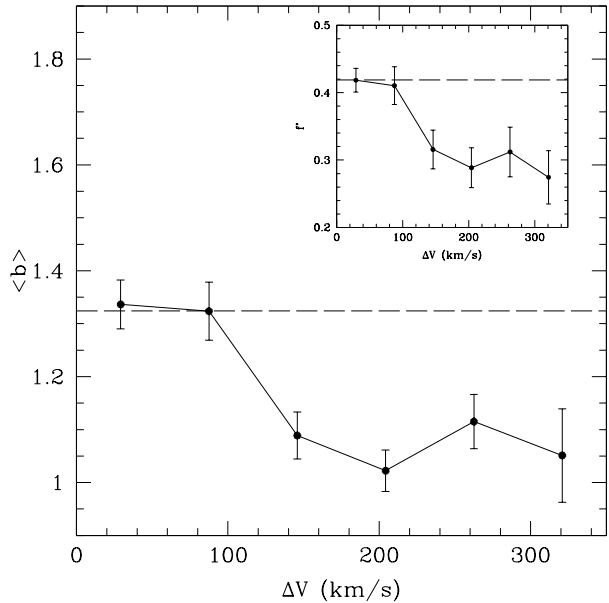}
\caption{Mean $b$ parameters estimated in projected distance bins for galaxies
in FGP catalog for three different maximum  relative radial
velocity limits: $\Delta V < 350 \ {\rm km/s}$ (solid line),
$\Delta V < 100 \ {\rm km/s}$ (long-dashed line) and
$\Delta V < 50 \ {\rm km/s}$ (dotted line).
The dashed horizontal line
represents the mean b parameter for the control sample.
The small box correspond to the fraction $f^\star$ of galaxies
with $b>\bar{b}$.
}
\label{orbital}
\end{figure}

In order to visualize how $<b>$   depends on the
combination of orbital parameters we have estimated  $b$ as
a function of $r_{\rm p}$
 by considering pairs
by  different velocity separations (Fig. ~\ref{orbital}).
 As it can be clearly seen
the signal increases as pairs with larger  $\Delta V$ are
excluded from the calculations. A similar behavior is observed for
the fractions ($f^{\star}$). Those are systems with
large relative distance observed with a small
projected separation.
From this figure we may infer that
as the galaxies orbit into each other or experience a close
encounter, the star formation activity
of the system increases. In spite of the uncertainty
in $\Delta V$ due to measurement errors, we note that  the
larger star formation activity for pairs with quoted values as small as
$\Delta V < 50 $ km/s.

\begin{figure}
\includegraphics[width=84mm]{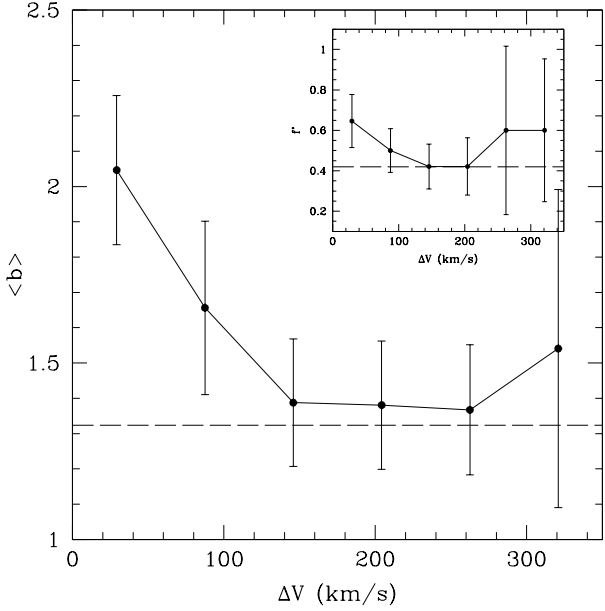}
\caption{Mean  $b$ parameters in  bins of relative radial velocity
for galaxies in  close  pairs  ($r_{\rm p} < 25$ kpc).
The dashed horizontal line represents the mean $b$
parameter for galaxies in the control sample.
The small box shows the fraction $f^{\star}$ of strong
star forming galaxies in the samples.
}
\label{br25}
\end{figure}

The most significant difference between the star formation activity
of galaxies in pairs and those of the control sample is obtained when
the analysis is restricted to pairs with $r_{\rm p} \leq 25 $ kpc.
This sub-sample  shows the strongest signal
of enhanced SF as illustrated in  Fig. ~\ref{br25}
where we  have plotted  the $b$ parameters as a function of
the relative radial velocity . The dashed lines depict the mean $<b>$ and $f^\star$
for the corresponding control sample. Confronting this figure with Fig. ~\ref{fig4B}
X we can conclude  that pairs with small $r_p$  have a statistically significant
increase of their star formation rate when restricted to
small relative velocities. In consequence, we define a subsample of close pairs
by imposing the restrictions:
$r_p < 25$ kpc and $\Delta V < 100 \  {\rm km/s}$, hereafter close GP (CGP)
sample.
This set of galaxies exhibits the  the   higher star formation efficiency.

\subsection{Dependence on Luminosity}

An analysis of the dependence of the SF activity on the luminosity of
the galaxies in pairs could help
to understand  how
SF activity is regulated between pair members.
Several observational works have found evidence that
in an interacting pair, the fainter galaxy seems to be
more affected (e.g. Donzelli \& Pastoriza 1997).
Recently, Colina et al. (2001) found that Ultra Luminous
Infrared Galaxies  could be related
to the merger of  structures with different luminosities
in the range $0.3-2L^{*}$.
 Arguments related to the stability properties
of the galaxies have been used to explain these observations.
 Numerical simulations show that disc galaxies
with a  small
or with  a lack of bulge component (e.g, Mihos \& Hernquist 1996;
 Dom\'{\i}nguez-Tenreiro,
Tissera \& S\'aiz 1998; Tissera et al. 2002) tend to be more
susceptible to the effects of tidal interactions which can trigger gas
inflows and star formation.
According to Tissera et al. (2002) these systems would be in
early stages of evolution where their bulges are being built up, and
consequently, on average, they would be smaller (or fainter).

We can use this large  galaxy pair catalog  for
 investigating this point
 on a  more firm statistical basis.
For this purpose we estimate the mean birth rate parameter
 in luminosity bins for
galaxies in the FGP catalog and those in  the control
sample. We then define the star formation excess   $\beta$
as the ratio between these two
$b$ parameters. {\it Hence, $\beta$ yields the excess of star formation
in galaxies in pairs with respect to any isolated galaxy in the field
with the same selection effects and redshift distribution.}
This point is crucial for the understanding of the following reasonings
since we are always assessing the effects of having a companion on the
SF activity by comparing to isolated galaxies with the same
redshift distribution. In this Section we also add the constrain of
luminosity range.

\begin{figure}
\includegraphics[width=84mm]{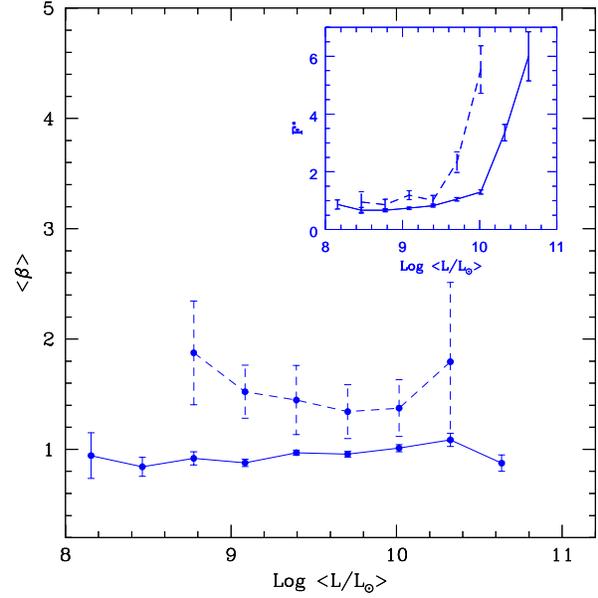}
\caption{ Mean star formation excess parameters
$\beta$ for
the FGPs (solid lines) and
CGPs  with projected distance $r_{\rm p} <25$ kpc
and  $\Delta V < 100 \ {\rm km/s}$ (dashed lines).
The small box shows the relative fraction $F^{\star}$ of strong
star formation.
}
\label{pbl-frac}
\end{figure}

In Fig.~\ref{pbl-frac} we show $<\beta>$ vs log $<L/L_{\odot}>$
for the FGP and the CGP samples ($r < 25$ kpc and
$\Delta V < 100 \  {\rm km/s}$).
As it can be seen the total
FGP sample shows the same level of SF activity than isolated galaxies in the
field ($\beta \approx 1$),
 while the CGP sample has a larger star formation enhancement with
respect to isolated galaxies ($\beta \ge 1.5 $).
In the small box, we show  the 
 fraction of galaxies ($F^{\star}$)
that are
experiencing larger SF activity than the corresponding mean in the control
sample, in each
luminosity interval, over the corresponding fractions of the control samples.
This calculation was carried out for both the complete and the close
galaxy pair samples.
 It can be appreciated from this box
a significant increase in the fraction of star formation galaxies
in pairs with respect to isolated ones as a function of
luminosity. This trend is washed out when mean b values are estimated since
there is a large percentage of non star-forming pairs.
Owing to the fact that the larger difference between the star formation
of galaxies in pairs and that of isolated ones is detected
for bright  galaxies, we have to consider the possible presence
of AGN as well as starburst associated to interactions.
Note that we are always measuring the SF excess respect to
isolated galaxies. In absolute  mean values, faint galaxies
have a higher $b$ parameter.

 \subsection{Pairs formed by similar and different luminosity
galaxies}

\begin{figure}
\includegraphics[width=84mm]{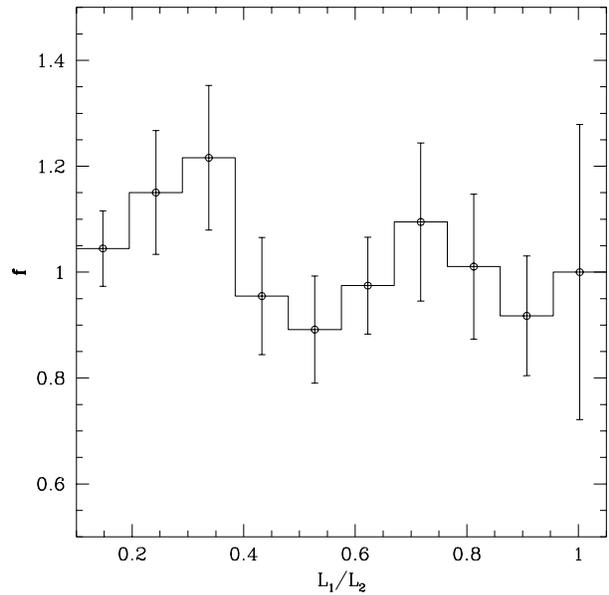}
\caption{Histogram of the luminosity ratio between the
bright and the faint galaxy member of pairs normalized to
that estimated  for the control sample.
}
\label{histol1l2}
\end{figure}

In this section  we  analyse FGPs according to
the relative luminosity of their member galaxies.
For FGPs and galaxies in the control sample,
 we estimate the ratio of luminosities  $L_2/L_1$ of the faint
over the bright member.  As shown in Fig. ~\ref{histol1l2},
we found that  FGPs are composed of galaxies with
relative luminosities similar to those of any isolated pair of galaxies
 in the field
(with the same redshift distribution),  except for a weak tendency
for  some excess of pairs of galaxies composed with
different luminosities.
The bootstrap error analysis indicates that this excess has a
statistical significance.
This is likely to be due to the effects of the interaction,
where it could be argued for truncated star formation owing to the
tidal perturbation of
the more massive companion.

We define two sub-samples  according to the
relative luminosity  of galaxies ($L_2, L_1$) in pairs.
We adopt   $L_2/L_1 =0.5$ as a threshold
 to split the FGP catalog  into two subsamples of dissimilar
($L_2/L_1 < 0.5$) and a similar
 ($L_2/L_1 \ge 0.5$) galaxy luminosities in pairs
(this choice divides the sample into subsamples with roughly the same number
of members).
 Assuming that all interacting
pairs might eventually  give origin to a merger event,
 this luminosity ratio  can be also interpreted as a threshold
 to split the data into major and minor merger candidate sub-samples.
 We then calculated the $<\beta>$ parameter as a function of luminosity
for these two sub-samples  restricted to the subsample of close galaxy pairs
defined by
$r_{\rm p} < 25 $ kpc and $\Delta V < 100 \ {\rm km/s}$.
While for the minor merger candidate sub-sample, we found no
excess, a mean $\beta \approx 2$ is detected for the major merger
candidate one. However, in both cases, we found no dependence on luminosity.

\begin{figure}
\includegraphics[width=84mm]{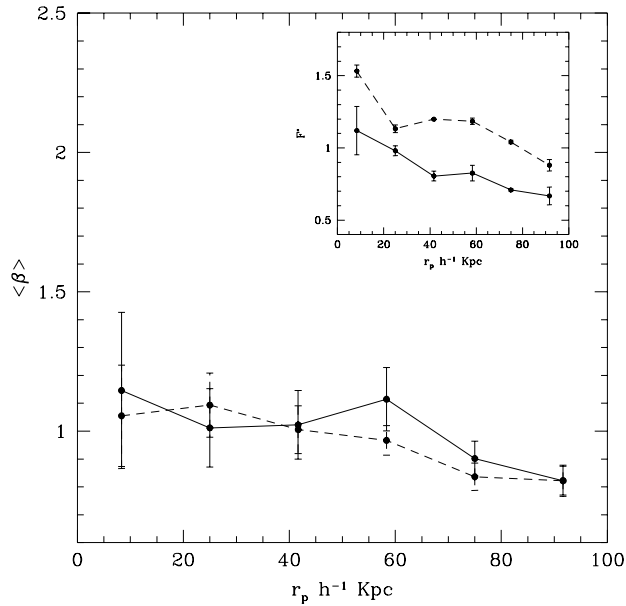}
\caption{Mean SF excess  parameter $\beta$
 versus projected distance for the bright (dashed lines)
 and the
faint (solid lines) members of CGPs
classified as minor merger candidates (i.e., pairs formed by
different luminosity galaxies, $L_1 >> L_2$).
The small box shows the relative fraction $F^{\star}$ for the
brighter (dashed lines) and the fainter (solid lines) member.
}
\label{minor}
\end{figure}

In order to shed light on the importance of galaxy luminosity and
the strength of the interaction
we have also investigated
the dependence of the SF excess on projected distance
for the fainter and brighter galaxy members in
minor merger candidates  ($L_2/L_1 \leq 0.5$).
For this purpose we have estimated the $<\beta>$ parameter
 as a function of $r_{\rm p}$ for
the faint and bright  components
of CGPs as  shown in
Fig. ~\ref{minor}.
It can be appreciated from this figure  the
similar behavior of the mean SF enhancement with respect to the control
sample of the two components.  However, when fractions are estimated for the
bright (dashed) and faint (solid) components we found that the former shows
a clear excess. At a given luminosity, brighter components of dissimilar
luminosity pairs have a larger probability to have
enhanced SF when compared to isolated galaxies and from larger
 projected distances.

\begin{figure}
\includegraphics[width=84mm]{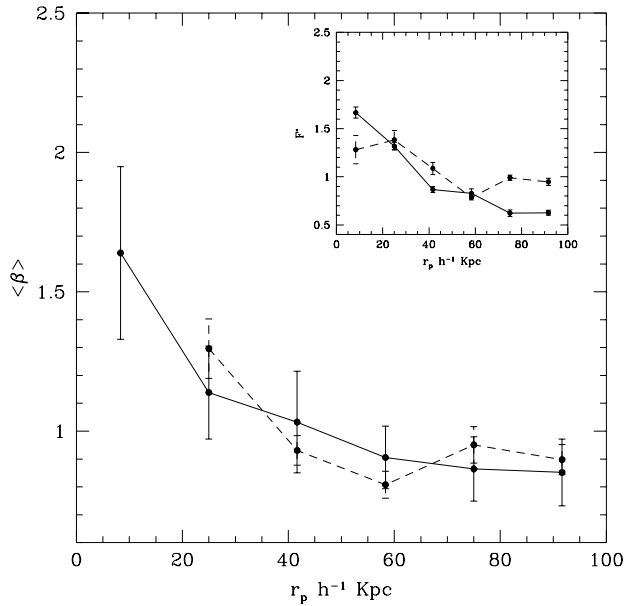}
\caption{Mean SF excess  parameter $\beta$
 versus projected distance for the bright (dashed lines)
 and the
faint (solid lines)  galaxy pairs
classified as major merger candidates (i.e., pairs formed by
different luminosity galaxies, $L_1 \sim L_2$).
The small box shows the relative fraction $F^{\star}$ for the
bright (dashed lines) and the faint (solid lines) member.
}
\label{major}
\end{figure}

As similar analysis was performed for the major merger candidate sub-sample.
In this case in order to classify them in bright and faint luminosity
pairs  we estimated the mean luminosity of the pair and
adopted a threshold of  log $(L_1 + L_2 )/(2L_{\odot}) = 9.5$.
For these two sets, we
estimated $<\beta>$ as a function of projected distance.
The results are plotted in Fig. ~\ref{major} from where it can be
appreciated that
 the two samples exhibit a similar trend in  mean SF enhancement
as a function of $r_{\rm p}$ which is found to be higher than
that of minor merger candidates.
The small box shows the ratio between the fraction of galaxies with $b > \bar{b}$
and  those corresponding to the control sample.
In this case, we found that both components of major merger candidates
have a similar probability to have enhanced SF activity. However,
we note that
the fractions are larger than those of the corresponding control sample
only for close encounters ($r_{\rm p} \leq 25 $ kpc)

This analysis suggests that the effects of interactions on the SF activity
is more important in the brighter components of pairs with different luminosities. For galaxies with
a similar luminosity companion, both objects
show a larger star formation activity than isolated galaxies.

It should be stressed that these results do not contradict
observations where the faint member of interacting
pairs has on average
a larger star formation activity since
these studies lack a proper  confrontation to
isolated galaxies (see Bergvall et al. 2003 for the first observational works that used this approach).

\section{Discussion and Conclusions}

We carried out a statistical analysis of 1258 galaxy pairs in the field
with $z \le 0.1$  selected from the  2dF public release data.
Our study is centred on the star formation enhancement of the pair members
with respect to
isolated galaxies in the field with the same redshift distribution and luminosities by analysis the dependence of the mean birth
rate paremeters and the fractions of star forming galaxies.
Particularly, the latter results to be an important
tool to unveil the effects of having a companion since
it is less affected by interpolers, non SF pairs, etc.

We can summarize our results in the following main conclusions.

1. We detect a significant
correlation between the star birth rate parameter $<b>$ and both,
projected spatial separation $r_p$
and  relative radial velocity $\Delta V$ for galaxy pairs.
For $r_p < 25 $kpc and $\Delta V < 100$ km/s we obtain a substantial star formation enhancement
with respect to the isolated control sample.
The $\Delta V$ dependence is less pronounced
although we find a systematic increase of star formation activity
for decreasing relative velocity.

2.  A significant percentage of galaxies ($\approx 45 \%$ )
which satisfy the
pair selection criteria but that have negligible star formation
has been found. This result suggests that it is not only
proximity that may be playing a role in triggering star formation but
the internal structure of the galaxies could also  be a crucial factor
as it has been found in numerical simulations.

3.  There is no an overall dependence of the
mean star formation enhancement
 $<\beta> $ on luminosity for galaxy pairs at $z \leq 0.10$.
A nearly constant value  $<\beta> \simeq 1.2  $  for the total FGP catalog
and
$<\beta> \simeq 2 $  for the close GP subsample have been measured.
However, the fractions of galaxies in pairs that have a higher
SF activity than the  corresponding averaged one  of isolated galaxies, show
a clear excess for the bright members.

4. We divide  the close pair sample in  minor ($L_2/L_1 < 0.5$) and major
($L_2/ L_1 > 0.5$) merger candidates
according to their relative luminosities.
We found that bright components in minor merger candidates show higher probability to have enhanced SF by tidal interactions than isolated
galaxies and from larger projected distances than the faint components.
 In the case of major mergers both components show
comparable star formation enhancements and with a
similar projected distance dependence.

\section*{Acknowledgments}

This work was partially supported by the
 Consejo Nacional de Investigaciones Cient\'{\i}ficas y T\'ecnicas,
Agencia de Promoci\'on de Ciencia y Tecnolog\'{\i}a,  Fundaci\'on Antorchas
 and Secretar\'{\i}a de Ciencia y
T\'ecnica de la Universidad Nacional de C\'ordoba.

\end{document}